\documentclass[11pt]{article}
\pdfoutput=1
\usepackage[english]{babel}

\usepackage{bm}%
\usepackage[utf8]{inputenc}
\usepackage{amsmath}
\usepackage{amsfonts}
\usepackage{graphicx}
\usepackage{graphicx,epstopdf}
\usepackage{geometry}
\usepackage{blindtext}
\usepackage{ulem}
\usepackage{hyperref}
\usepackage{amsthm}
\usepackage{authblk}
\newtheorem{theorem*}{Theorem}[]
\usepackage{enumerate}
\usepackage[utf8]{inputenc}

\hypersetup{
    colorlinks=true,
    filecolor=magenta,      
    urlcolor=red,
    citecolor=red,
    }

\title{Optimal Locomotion for Limbless Crawlers}
\author[1]{Sreejith Santhosh}
\author[1]{Mattia Serra \thanks{mserra@ucsd.edu}}
\affil[1]{Department of Physics, University of California San Diego}

\begin{document}

\maketitle
\begin{abstract}
Limbless crawling is ubiquitous in biology, from cells to organisms. We develop and analyze a model for the dynamics of one-dimensional elastic crawlers, subject to active stress and deformation-dependent friction with the substrate. We find that the optimal active stress distribution that maximizes the crawler's center of mass displacement given a fixed amount of energy input is a traveling wave. This theoretical optimum corresponds to peristalsis-like extension-contraction waves observed in biological organisms, possibly explaining the prevalence of peristalsis as a convergent gait across species. Our theory elucidates key observations in biological systems connecting the anchoring phase of a crawler to the retrograde and prograde distinction seen in peristaltic waves among various organisms. Using our optimal gait solution, we derive a scaling relation between the crawling speed and body mass, explaining experiments on earthworms with three orders of magnitude body mass variations. Our results offer insights and tools for optimal bioinspired crawling robots design with finite battery capacity.  
\end{abstract}

\section{Introduction}
Limbless crawling is a form of locomotion performed by various organisms, including planaria, leeches, nemertea, aplysia, snails, chitons, earthworms and larvae \cite{alexander1992exploring,Gray68,LowerAnimals,LISSMANN,Trueman}. Crawling motion is also observed in motile cells during embryonic development, cancer metastasis, healing, and inflammation \cite{CellCrawling}. Limbless crawlers move using muscular contraction-extension waves across their body, and the interaction with the substrate \cite{LowerAnimals}. Caterpillars and millipedes also perform limbless crawling, with legs functioning as anchors responding to the body's extension contraction \cite{Gray68}. This type of locomotion is helpful in harsh terrains and cramped spaces,  generating interest in the biomimetic-robots community for applications in search and rescue, endoscopy and burrowing \cite{8324521,boxerbaum,Fang2015PhaseCA,Daltorio_2013,Nemitz,MenciassiSoft,EndoScope}. 
\\
\\
Out of the infinitely many possible gaits for limbless crawlers, a natural question is why these systems exhibit a particular one, made of extension-contraction waves \cite{alexander1992exploring,Trueman}. The existence of peristaltic waves as a convergent evolutionary trait across various organisms points to a strong selection pressure. To analyze this question, we derive a theoretical continuum framework for the dynamics of a limbless crawler. Our work allows for a continuum description and analysis of crawling organisms and soft robots, adding to the existing literature modeling crawlers as metameric (or discrete) systems \cite{Fang2015PhaseCA,OptimalGait,Tanka}. We analytically derive optimal locomotion strategies that are the most energy-efficient. In addition to addressing the question of constrained evolutionary optimization, our findings are relevant for biomimetic crawling robots, as one fundamental constraint for mobile robots is the finite battery capacity \cite{AKhandari2020,OptimalGait}. Several studies have dealt with the optimization of crawler gaits with prior assumed functional forms of actuation or have done so in models with finite degrees of freedom \cite{Fang2015PhaseCA,OptimalGait,KELLER1983417}. 
\\ \\
We provide the exact form of the optimal extension-contraction waves and the active stress distribution required to achieve it. The energy-efficient nature of these wave solutions may explain why we observe such gaits as convergent evolutionary traits, and aid optimal soft robots design. Our model also explains the difference between crawlers with prograde and retrograde peristaltic waves, where prograde waves move from tail to head as in \textcolor{black}{\textit{Nereis diversicolor}}, and retrograde waves move from head to tail as in earthworms \cite{NerisGray,EarthwormGray}. Our analytical result for the velocity of the peristaltic wave precisely relates the anchoring behavior of the crawler with the prograde or retrograde motion. \textcolor{black}{Using our optimal gait solution, we find a scaling law between the crawling speed and body mass, and compare it with experimental data on earthworms with three orders of magnitude body mass variations \cite{Quillin}.}

\section{Methods}                                                          

\begin{figure}[h]
\centering
\includegraphics[width=\textwidth]{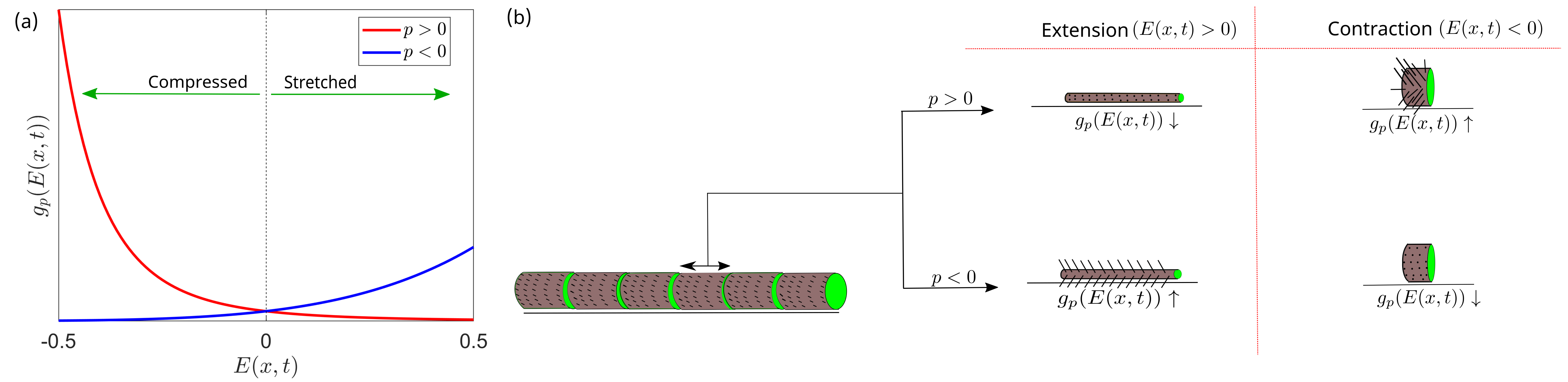}
\caption{(a) Stretch-dependent friction coefficient $g_p(E) = \frac{1}{(1+E(x,t))^p}$. (b) The ``p-model of friction" captures two different anchoring phases with $p>0$ and $p<0$. 
}
\label{fig:bristlesWorm}
\end{figure}

We model crawlers as an active one-dimensional continuum that can move in one dimension \cite{alexander1992exploring,Gray68}. To move forward, the crawler exerts a backward force on the underlying substrate via the formation of anchors across its body. Anchors' modulation allows both holding and moving body segments on the substrate. Such modulation arises by coupling anchoring to stretch along the body. We use the ``p-model" \cite{Fang2015PhaseCA,OptimalGait} to describe the simplest substrate anchoring behavior, in which the strain controls the anchoring strength. The frictional force is viscous and satisfies the relation
\begin{equation}\label{eq:friction mod}
    f(x,t) = -\gamma g_p(E(x,t))v(x,t), \quad g_p(E) = \frac{1}{(1+E)^p},
\end{equation}
where $x$ is the 1-D coordinate along the body, $t$ is the time, $E(x,t)$ is the strain, $v(x,t)$ is the velocity, $p\in (-\infty,\infty)$,  and $\gamma$ is the coefficient of substrate friction. For $p=0$, friction reduces to the Newtonian model of viscous drag. For $ p \to \infty$, this friction model produces an idealized condition in which no force opposes slip when the body is extended ($E>0$), and it can withstand any tangential force when it is contracted ($E<0$), vice-versa for $p\to -\infty$. This friction model is inspired by earthworm-setae (Fig. \ref{fig:bristlesWorm}), which protrude when the body is axially contracted resulting in a higher resistance, and has been used \cite{Kano,OptimalGait,Fang2015PhaseCA} to describe metameric robot kinematics, which exhibits earthworm-setae-like frictional dynamics.
Different species have different anchoring behavior which are usually modulated by $E(x,t)$ \cite{Trueman}. We generalize $p\in (-\infty,\infty)$ to accommodate these wide varieties of anchoring behaviour, even though eq. \eqref{eq:friction mod} was originally derived for earthworm-setae friction dynamics, which has $p>0$.

The Equation of Motion (EOM) of a 1-D continuum where inertia forces are negligible follows from the linear momentum balance 
\begin{equation}\label{eq:ViscousEulerianEOM}
    f(x,t)+\frac{\partial m(x,t)}{\partial x}+\frac{\partial(k(E(x,t)-1))}{\partial x}=0,
\end{equation}
where $f(x,t)$ is the frictional force \eqref{eq:friction mod}, $m(x,t)$ is the active stress and $k(E(x,t)-1)$ is elastic stress generated across the body where $k$ is the constant elastic modulus. Because the endpoints of the crawler move in a priori unknown locations, the Eulerian frame is not suitable, and we rewrite \eqref{eq:ViscousEulerianEOM} in Lagrangian coordinates, choosing the initial un-stretched state as the reference configuration. We denote by $s\in [0,L]$ any initial position along the body with initial length $L$, and by $\zeta(s,t) = s + \int_0^t v(\zeta(s,\tau),\tau)d\tau$ its corresponding time-$t$ position. The linear momentum balance (\ref{eq:ViscousEulerianEOM}) in the Lagrangian frame (see \ref{TransformEulerLagragian}) is given by,
\begin{equation}\label{eq:FinalLagEOM}
     \gamma \zeta_t\zeta_s^{a}= m_s+k\zeta_{ss},\quad a=1-p,
\end{equation}
where $\zeta_{(\cdot)} = \partial_{(\cdot)}\zeta$ and $m_{(\cdot)} = \partial_{(\cdot)}m$. 
\textcolor{black}{The force per unit body length in the reference configuration ($-\gamma \zeta_t\zeta_s^{a}$) accounts for both the change in the frictional coefficient ($-\gamma \zeta_t\zeta_s^{-p}$) and the change in substrate contact length due to strain ($-\gamma \zeta_t\zeta_s$).} $m_s$ and $k \zeta_{ss}$ are the active and elastic forces arising from spatial variations of active stress and elastic strain.  
\begin{figure}[h]
\centering
\includegraphics[width=\textwidth]{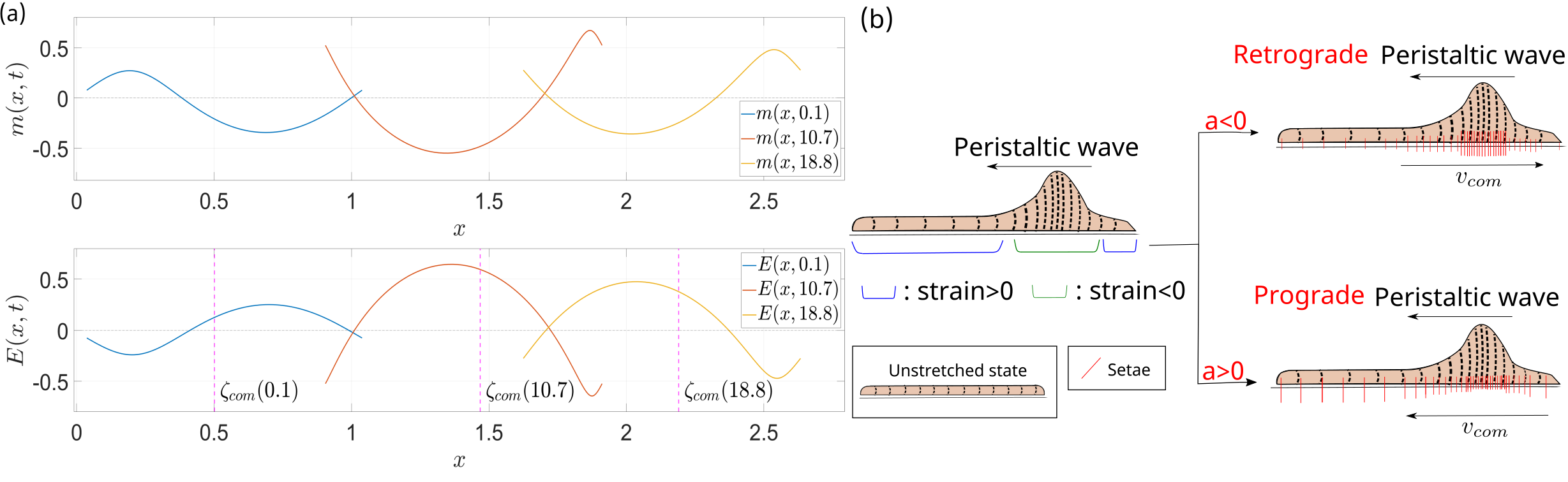}
\caption{(a) Snapshots of active stress $m(x,t)$, strain $E(x,t)$ and $\zeta^{com}(x,t)$ for a crawler performing optimal locomotion. The time evolution of the dynamics is available in \href{https://drive.google.com/file/d/1hZ-Bkeo4KVUSvF0vHTC-Sr1uDdyuCgab/view?usp=sharing}{[Movie]}. The top panel is the active stress $m(s,t)$ across the crawler . Parameters for the simulation are $a= -1,\text{ } \gamma = 1,\text{ } k=1,\text{ } H= 10,\text{ } L=1,\text{ } T = 20,\text{ } \chi = 0.4,\text{ } a_1 = 1,\text{ }a_{n\neq1}=0,\text{ }b_n = 0 \text{ }\forall \text{ }n\in \mathbb{N}^+$. (b) The anchoring behavior of the crawler determines if the peristaltic motion is either retrograde ($a<0$) or prograde ($a>0$). \textcolor{black}{Where the crawler is extended, two effects increase the anchoring to the substrate. First, the stretch-dependent variation in frictional coefficient $\zeta_s^{-p}$ increases. Second, the increase of contact length with the substrate due to stretch-dependent density variation $\zeta_s$.
The dashed black lines are equally spaced in the undeformed body configuration and aid the visualization of the mass density variation during motion.}}
\label{fig:ForcesAndOptimalMotion}
\end{figure}

\section{Results}
The crawling system described by Eq. (\ref{eq:FinalLagEOM}) moves by controlling the active stress generated across its body, typically dictated by the nervous system in organisms \cite{pehlevan2016integrative,paoletti2014proprioceptive} and control units in robots \cite{umedachi,rus2015design}. Various mechanisms can generate active stress, ranging from actomyosin filaments in biology to actuators in robotics. Given the infinitely large space of strategies for $m(s,t)$, we seek the energy-optimal distribution of $m(s,t)$ to perform locomotion in limbless crawling systems. We define optimality as moving the farthest using a fixed energy reservoir. We find the optimal solution by solving an infinite-dimensional variational problem (\ref{sec:FrameVariational}). Given the nonlinear nature of the EOM, we use perturbation series methods and solve the variational problem (\ref{PerturbAnalysis} and \ref{sec:SolvingVariationalProblem}) to get the following result.

\begin{theorem*}\label{theorem}
Consider a one-dimensional continuum whose dynamics satisfies eq. (\ref{eq:FinalLagEOM}). The optimal active stress distribution $m(s,t)$ that achieves maximum center of mass displacement by time $T$ given a fixed amount of total energy available, in the small deformation limit, is given by the extrema of the following functional,
\begin{multline}
  S = \underbrace{\frac{1}{L}\int_{0}^{T}\int_{0}^{L}\zeta_t dsdt}_{\delta\zeta^{com}}+\underbrace{\alpha(\int_{0}^{T}\int_{0}^{L}m_s\zeta_tdsdt-V)}_{\text{energy input constraint}}+\underbrace{\int_{0}^{T}\int_{0}^{L}\lambda(s,t)(-\gamma \zeta_t\zeta_s^{a}+m_s+k\zeta_{ss})dsdt}_{EOM}\\+\underbrace{\int_{0}^{T}\theta_1(t)(m(0,t)+k(\zeta_s(0,t)-1))dt +\int_{0}^{T}\theta_2(t)(m(L,t)+k(\zeta_s(L,t)-1))dt.}_{\text{stress-free boundary conditions}}
\end{multline}
Here $\alpha$, $\lambda(s,t)$, $\theta_1(t)$ and $\theta_2(t)$ are Lagrange multipliers and $V$ is the total energy reservoir. The optimal $m(s,t)$ in Fourier basis\footnote{The general form of the optimal $m(s,t)$ is given by Eq.  (\ref{eq:Finalmst}).} is given by
\begin{multline}
    m(s,t) = \chi(\gamma(\sum_{n=1}^{n=\infty}a_n\frac{(\cos(q_nct)-\cos(q_n(s+ct)))}{q_n}+b_n\frac{(\sin(q_n(s+ct))-\sin(q_nct))}{q_n})\\-\frac{k}{c}(\sum_{n=1}^{n=\infty}a_n(\sin(q_n(s+ct))-\sin(q_ns))+b_n(\cos(q_n(s+ct))-\cos(q_ns))))+O(\chi^2),
\end{multline}
where $0 < \chi\ll 1$ is a perturbation parameter controlling the stretching of the system $\zeta_s(s,t)=1+ \sum_{i=1}^{\infty}\chi^i p^i(s,t)$, $c=\frac{a}{L\gamma\alpha}$, $q_n = \frac{2\pi n}{L},\text{ }n\in \mathbb{N}^+$ and $\alpha$ is determined by the energy equation
\begin{equation}\label{eq:EnergyAlpha}
H = 2L\gamma T\Delta_2+\frac{2L^3\gamma^2k\alpha^2}{a^2}(\Delta_2-\sum_{n=1}^{n=\infty}\cos(\frac{2n\pi Ta}{L^2\gamma\alpha})\frac{(a_n^2+b_n^2)}{4}), 
\end{equation}
where $V = \chi^2H+O(\chi^3)$, $\chi^2\Delta_2$ is the initial kinetic energy, $T$ is the total motion time, and $a_n$ and $b_n$ are Fourier coefficients. In the case of multiple roots $\{\alpha_i\}$, we select $\alpha$ such that $\delta\zeta^{com}(\alpha)$ is maximum, where the total center of mass displacement is given by
\begin{equation}\label{eq:xComTheorem}
    \delta \zeta^{com}(\alpha)= \chi^2(-2L\gamma\alpha T\Delta_2+\frac{L^3\gamma^2\alpha^2}{4\pi a}\sum_{n=1}^{n=\infty}\sin(\frac{2n\pi Ta}{L^2\gamma\alpha})(\frac{a_n^2+b_n^2}{n}))+O(\chi^3).
\end{equation}
The strain experienced by the system under the optimal $m(s,t)$ is given by
\begin{equation}\label{eq:StrainCrawler}
    \zeta_s(s,t)-1 = \frac{\chi}{c}(\sum_{n=1}^{n=\infty} a_n(\sin(q_n(s+ct))-\sin(q_ns))+b_n(\cos(q_n(s+ct))-\cos(q_ns))+O(\chi^2).
\end{equation}
\end{theorem*}

\begin{proof}
See supplementary information \ref{sec:FrameVariational}-\ref{sec:SpectralAnalysis}.
\end{proof}
In Fig. \ref{fig:zetaCOM}, we check the validity of our perturbation series solution by comparing the motion of the center of mass from eq.  \eqref{eq:xComTheorem} with the one from the full nonlinear system \eqref{eq:FinalLagEOM} for various values of $\chi$, and find increasing accuracy for decreasing $\chi$.

\subsection{Comparison with biology of crawling organisms}

The optimal traveling-wave solutions derived in Theorem \ref{theorem} is observed in limbless crawling organisms. This particular gait is known as peristaltic locomotion, which consists of extension-contraction waves along the body. There are two distinct peristaltic waves observed in nature: retrograde and prograde waves. We explain this distinction using the analytical result for the velocity of the traveling wave given by $c=\frac{a}{L\gamma\alpha}$, as defined in Theorem \ref{theorem}. \textcolor{black}{From \eqref{eq:xComTheorem}, the center of mass performs both directed motion $-2L\gamma \alpha T\Delta_2$ and oscillatory motion $\frac{L^3\gamma^2\alpha^2}{4\pi a}\sum_{n=1}^{n=\infty}\sin(\frac{2n\pi Ta}{L^2\gamma\alpha})(\frac{a_n^2+b_n^2}{n}))$. If $\alpha<0$, the crawler performs directed motion in the positive x-axis.} Depending on the value of $a$, peristaltic waves can either be retrograde or prograde. $a>0$ results in prograde waves and $a<0$ results in retrograde waves (Fig. \ref{fig:ForcesAndOptimalMotion}b). \textcolor{black}{The same result holds for $\alpha>0$ but for motion of the center of mass in the negative x-axis. } Using this result, we can relate the anchoring phase ($p$) to the retrograde and prograde motion of the peristaltic wave for different organisms. 
\\
\\
\textcolor{black}{The stretch-dependent frictional force per unit length is $\propto \zeta_s^{-a}$ (\ref{eq:FinalLagEOM}) and it captures both the effect of stretch-dependent variation in frictional coefficient $\zeta_s^{-p}$ and the mass density variation with stretching $\zeta_s$. The peristaltic wave is retrograde for an earthworm, implying $a<0$ and therefore $p>1$. This is consistent with biological observations where the substrate friction increases where the earthworm's body is contracted (or denser). 
Similarly, for worms like \textit{Nereis diversicolor}, where the peristaltic wave is prograde, we infer that their anchoring behavior falls in the regime of $p<1$. If $p<1$ the body segment anchors (increased friction) during extension and performs free slip (reduced friction) during contractions. This is consistent with the biology of \textit{Nereis diversicolor}, where the anchors are formed in the extended part of the body and let the body slide in contracted regions \cite{NerisGray}. We can extend this finding to snails and slugs, which perform prograde peristaltic locomotion \cite{alexander1992exploring,Trueman}. The mucus on the foot of snails and slugs acts as a set of solid anchors in the extended part of the body and lets the body slide in contracted regions. Therefore even in organisms that perform peristaltic locomotion without using seate-like protrusions to generate frictional anisotropy, our results provide insights into the underlying anchoring behavior.}

\subsection{Scaling of crawling locomotion}

\begin{figure}[h]
\centering
\includegraphics[width=0.5\textwidth]{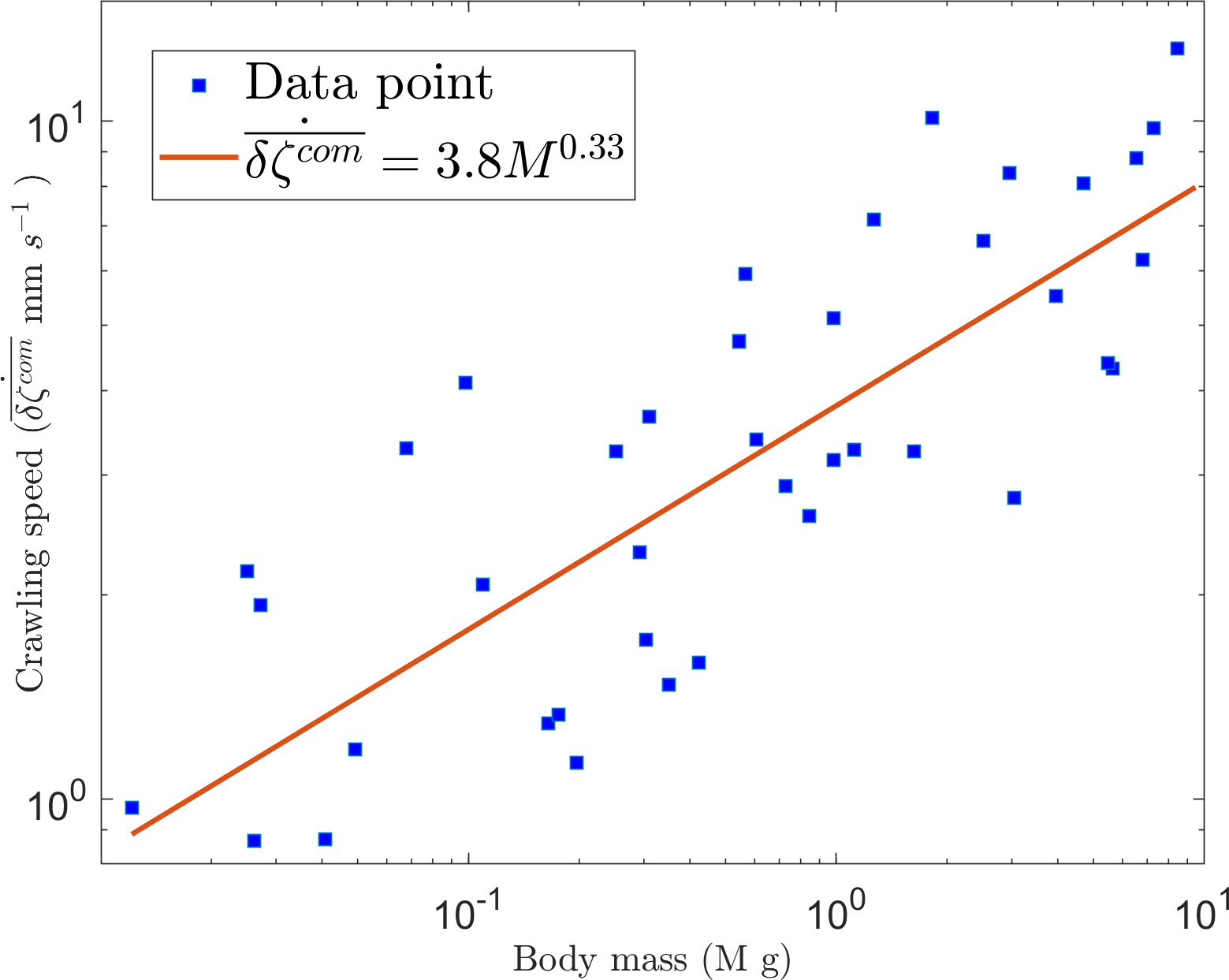}
\caption{ Experimental data points (blue square dots) and experimental fit curve (red) of crawling speed $\dot{\overline{{\delta\zeta^{com}}}}$ as a function of body mass $M$, reproduced from \cite{Quillin}. The crawling speed $\dot{\overline{{\delta\zeta^{com}}}}$ is fit to a scaling curve $\dot{\overline{{\delta\zeta^{com}}}} = A_0 M^{d}$ with $A_0 = 3.8$ and $d=0.33$ (correlation coefficient $r^2 = 0.66$ and P-value $<0.001$), consistent with our analytical prediction \eqref{eq:ScalingNormVel}.}
\label{fig:ScalingCrawlingSpeed}
\end{figure}

We use our analytical results in Theorem \ref{theorem} to study the scaling behavior of the crawling gait with the body size. From \eqref{eq:xComTheorem}, the average crawling speed $U$ normalized to the body length $L$ over a time period $T$ (see \ref{sec:ScalingRelations}) is given by
\begin{equation}\label{eq:MainCrawlingSpeed}
    U = \frac{\delta\zeta^{com}}{LT} = -2a\nu A^2,
\end{equation}
where $A^2 = \frac{\chi^2\Delta_2}{c^2}$ and $\nu$ is the stride frequency. \textcolor{black}{We assume that the modulation $(a=1-p)$ of friction by strain \eqref{eq:friction mod} is independent of the body size}. Since $A$ scales with the strain wave amplitude $\frac{\chi a_n}{c}$ and $\frac{\chi b_n}{c}$, its scaling behavior is constrained by the maximum tensile stress, which is upper bounded the failure stress of the body material. Therefore, the maximum sustainable strain of the elastic crawler body depends on material properties and not on body size, hence 
\begin{equation}\label{eq:MainScalingA}
    A\propto M^{0},
\end{equation}
where $M$ is the body mass. Our argument is in agreement with the experimental data on earthworm \textit{Lumbricus Terrestris} (cf. Fig. 6 in \cite{Quillin}). Using \eqref{eq:MainScalingA} and \eqref{eq:MainCrawlingSpeed}, we then find that $U\propto \nu$. To find the scaling behavior of $\nu$, we use the metabolic cost of transport (COT), which is the energy required to move a unit mass per unit distance \cite{alexander1992exploring}. The COT for the gait described in Theorem \ref{theorem}, using isometric growth relations as reported in \cite{QuillinIsometry}, is given by (see \ref{sec:ScalingRelations}) 
\begin{equation}\label{eq:MainCOTscaling}
    \text{COT} = \frac{V}{M\delta\zeta^{com}} \propto \nu. 
\end{equation}
 We interpret the stride frequency of the earthworm to be upper bounded by the COT \eqref{eq:MainCOTscaling} and lower bounded by the crawling speed $U$ \eqref{eq:MainCrawlingSpeed}. \textcolor{black}{To understand this trade-off, assume the scaling $\nu\propto M^{d}$ for $d>0$. In this case, the gait becomes energetically infeasible for larger $M$. Similarly, for $d<0$, the normalized crawling speed $U$ would be undesirable (low) for large $M$. For a crawler that optimizes for both $U$ and COT, the optimal scaling for $\nu$ would be $\nu\propto M^{0}$, consistent with the experimental results for the earthworm \cite{Quillin}. This helps maintain a constant gait across a wide range of body sizes without compromising on either $U$ or COT, which is especially important for earthworms, as they grow by three orders of magnitude \cite{QuillinIsometry}. Our efficiency-based argument agrees with biological explanations \cite{Gray68}, arguing that neural oscillators, which are independent of body size, set the stride frequency $\nu$ in earthworms.} Using the scaling relation for $\nu$ and $A$ in \eqref{eq:MainCrawlingSpeed}, we obtain  
\begin{equation}\label{eq:ScalingNormVel}
    U\propto M^{0},\quad \dot{\overline{{\delta\zeta^{com}}}}=LU=\frac{\delta\zeta^{com}}{T}\propto M^{1/3},
\end{equation}
 resulting in the crawling speed $\dot{\overline{{\delta\zeta^{com}}}}\propto M^{1/3}$. Figure \ref{fig:ScalingCrawlingSpeed} shows that our scaling relation for crawling speed explains the experimental data on earthworms \cite{Quillin}, \textcolor{black}{which measured the crawling speed $\dot{\overline{\delta\zeta^{com}}}$ for $N =41$ earthworms with a body mass $M$ ranging from 0.012-8.5 g.} 
 
\section{Discussions}
The convergent evolution of peristaltic locomotion in taxonomically diverse limbless crawling organisms suggests a selection pressure that operates across different physical and physiological constraints \cite{MaynardOptimization}. We hypothesize that the selection pressure maximizes the organism's displacement constrained to a fixed energy reserve. We devise a model for limbless crawler dynamics and solve the constrained optimization criterion analytically. We find that the optimal locomotion strategy is consistent with the peristaltic gait observed in nature. The optimal active stress that maximizes the displacement of the center of mass of the crawler given a fixed amount of energy input is a traveling wave across the crawler body. \textcolor{black}{Our findings provide a precise connection between the anchoring phase of a crawler to the retrograde and prograde distinction seen in peristaltic waves among various organisms \cite{Gray68,alexander1992exploring}. Finally, we derive a scaling relation between the crawling speed and body mass, explaining experimental data performed on $N = 41$ earthworms with three orders of magnitude body mass variations.}
\textcolor{black}{Our framework is limited to axial deformations of crawlers and does not account for gaits as lateral undulation in snakes and out of plane motion in inchworms \cite{alexander1992exploring}. These undulatory gaits can be thought of as a transition from axial to bending deformations caused by buckling of the crawler \cite{mahadevan2004biomimetic,LateralBendingDaniel}.} 
Our results apply to the analysis and control of crawling continuum robots \cite{kolachalama2020continuum}, with relevance from medical to industrial applications \cite{reviewcontinuumAct,robinson1999continuum} because of their high maneuverability and compliant material response \cite{rus2015design}. \\
\\

\textbf{Acknowledgments} We acknowledge Kim Quillin for providing the experimental data used in Figure 3. We also acknowledge Fabio Giardina, Nick Gravish, Olga Dudko and L. Mahadevan for their feedback on the manuscript.  

\newpage

\appendix

\setcounter{figure}{0}
\setcounter{equation}{0}
\renewcommand{\thefigure}{S\arabic{figure}}
\renewcommand{\theequation}{S\arabic{equation}}
\renewcommand{\thesection}{S\arabic{section}}
\section*{Supplementary Information}
\setcounter{section}{+1}
\subsection{EOM in Lagrangian frame}\label{TransformEulerLagragian}
Here we describe the dynamics of the limbless crawler (\ref{eq:ViscousEulerianEOM}) in the Lagrangian frame. The velocity of a body segment is given by
\begin{equation}\label{eq:LagEOMwithout}
    v(x=\zeta(s,t),t) =\frac{d\zeta(s,t)}{dt}=\partial_t \zeta(s,t).
\end{equation}
Strain is given by $E(\zeta(s,t),t) = \frac{\partial(\zeta(s,t)-s)}{\partial s}=\zeta_s(s,t)-1$,  elastic stress is given by $k(\zeta_s(s,t)-1)$, the active stress is given by $m(\zeta(s,t),t)=m(s,t)$ and the frictional force is given by $f(\zeta(s,t),t)=f(s,t) = -\gamma \frac{1}{\zeta_s(s,t)^p}\zeta_t(s,t)$. We can therefore rewrite (\ref{eq:ViscousEulerianEOM}) as
\begin{equation*}
    \gamma \zeta_t(s,t)\zeta_s(s,t)^{a}= m_s(s,t)+k\zeta_{ss}(s,t),\ \text{    } a=1-p.
\end{equation*}

\subsection{Perturbation series analysis}\label{PerturbAnalysis}
Since (\ref{eq:FinalLagEOM}) is non-linear and analytically intractable, we use perturbative methods to derive analytical results. We assume small deformation for the continuum body, which lets us write strain as a perturbation series
\begin{equation}\label{eqn:zetasInitDef}
   \zeta_s-1 = \sum_{i=1}^{\infty}\chi^i p^i(s,t),
\end{equation}
where $\lvert\chi\rvert<< 1$ is the perturbation parameter. Using (\ref{eqn:zetasInitDef}) we define a perturbation series for $\zeta(s,t)$,
\begin{equation}\label{eq:ZetaPertSeries}
    \zeta(s,t) = s+h(t)+\sum_{i=1}^{\infty}\chi^i\int_{0}^{s}p^i(s',t)ds'= s+h(t)+\sum_{i=1}^{\infty}\chi^i\beta^i(s,t),
\end{equation}
where $\int_{0}^{s}p^i(s',t)ds'=\beta^i(s,t)$. We also expand $m(s,t)$ in terms of a perturbation series as
\begin{equation}\label{eq:ActiveStressPertSeries}
  m(s,t) = \Gamma^0(s,t)+ \sum_{i=1}^{\infty}\chi^i\Gamma^i(s,t).
\end{equation}
The stress-free boundary condition implies $m(s,t)+k(\zeta_s(s,t)-1)=0$ at $s=0,L$. We analyze this condition using perturbation series and separate out relation between $m(s,t)$ and $\zeta_s(s,t)$ at different orders of $\chi$ as follows
\begin{equation*}
    O(\chi^0) \rightarrow \Gamma^0(0,t)=\Gamma^0(L,t)=0,
\end{equation*}
\begin{equation}\label{eq:SFBCi}
    O(\chi^i)\rightarrow \Gamma^i(0,t)+k\beta^i_s(0,t)=\Gamma^i(L,t)+k\beta^i_s(L,t)=0;\text{  }0<i\in \mathbb{N}^+.
\end{equation}

\subsection{Crawler dynamics at different orders in $\chi$}
We expand (\ref{eq:FinalLagEOM}) as a perturbation series using (\ref{eq:ActiveStressPertSeries}) and (\ref{eq:ZetaPertSeries}), and analyze the equation of motion at different orders of $\chi$.

\subsubsection{$O(\chi^0)$}
At zeroth order in $\chi$, (\ref{eq:FinalLagEOM}) is given by
\begin{equation}\label{eq:zeroordereqn}
-\gamma h_{t}(t)+\Gamma^0_s(s,t) = 0.
\end{equation}
Since $\Gamma^0_s(s,t)= \Gamma^0_s(t) = \gamma h_{t}(t)$, we can write $\Gamma^{0}(t) = s\gamma h_t(t)+c(t)$. From (\ref{eq:SFBCi}) we know that $\Gamma^0(s=0/L,t)=0$ which implies $c(t)=h_t(t)= 0$ and $\Gamma^0_s(s,t)= 0$. Therefore, at the zeroth order the crawler doesn't move and the gradients of active stress is zero.

\subsubsection{$O(\chi^1)$}
Equation (\ref{eq:FinalLagEOM}) at  $O(\chi^1)$ is given by
\begin{equation}\label{eq:firstordereqn}
-\gamma \beta^1_{t}(s,t)+\Gamma^1_s(s,t)+k\beta^1_{ss}(s,t)= 0.
\end{equation}
To analyze this equation we take a spatial average $<\cdot>_L = \frac{1}{L}\int_{0}^{L}(\cdot)ds$ to obtain 
\begin{equation}\label{eq:FirstOrderEOM}
-<\gamma \beta^1_{t}(s,t)>_L+<\Gamma^1_s(s,t)+k\beta^1_{ss}(s,t)>_L = 0.
\end{equation}
Substituting (\ref{eq:SFBCi}) in (\ref{eq:FirstOrderEOM}), we obtain
\begin{equation}\label{eqn:beta1evolv}
    <\beta^1_{t}(s,t)>_L = 0.
\end{equation}
The center of mass (COM) velocity is given by $\zeta^{com}_t(t) = \frac{1}{M}\int_{0}^{L}\rho_0\zeta_t(s,t)ds$, where $\rho_0 = \frac{M}{L}$ and $M$ is the total mass of the crawler. We can thus write $\zeta^{com}_t(t)$ as a perturbation series
\begin{equation*}
    \zeta^{com}_t(t) = \sum_{i=1}^{\infty}\chi^i<\beta_t^i(s,t)>_L,
    \label{eq:COMvel}
\end{equation*}
and observe that $<\beta^1_{t}(s,t)>_L = 0$ implies no COM motion at $O(\chi^1)$.

\subsubsection{$O(\chi^2)$}
Equation (\ref{eq:FinalLagEOM}) at  $O(\chi^2)$ is given by
\begin{equation}\label{eq:secondordereqn}
    -\gamma \beta^2_{t}(s,t)-a\gamma \beta^1_{t}(s,t)\beta^1_{s}(s,t)+\Gamma^2_s(s,t)+k\beta^2_{ss}(s,t) = 0.
\end{equation}
As before, we take a spatial average of (\ref{eq:secondordereqn}) and use (\ref{eq:SFBCi}) to obtain 
\begin{equation}\label{eq:O2Relation}
    <\beta^2_{t}(s,t)>_L = -a <\beta^1_{t}(s,t)\beta^1_{s}(s,t)>_L.
\end{equation}
Therefore, the $O(\chi^2)$ term of $\zeta^{com}_t(t)$ is the leading order term that contributes to the COM motion.

\subsection{Variational Problem}\label{sec:FrameVariational}

We seek the optimal $m(s,t)$ as the solution of infinite-dimensional variational problem \cite{CalcOfVariations} which maximizes the displacement of the COM with a given amount of energy supplied to the system. We constraint the space of possible functions by enforcing the EOM and stress-free boundary conditions with Lagrange multipliers, as described by the augmented functional 
\begin{multline}
  S(\zeta_t,\zeta_s,\zeta_{ss},m_s) = \underbrace{\frac{1}{L}\int_{0}^{T}\int_{0}^{L}\zeta_t(s,t)dsdt}_{\delta\zeta^{com}}+\underbrace{\alpha(\int_{0}^{T}\int_{0}^{L}m_s(s,t)\zeta_t(s,t)dsdt-V)}_{\text{energy input constraint}}\\+\underbrace{\int_{0}^{T}\int_{0}^{L}\lambda(s,t)(-\gamma \zeta_t\zeta_s^{a}+m_s+k\zeta_{ss})dsdt}_{EOM}+\underbrace{\int_{0}^{T}\theta_1(t)(m(0,t)+k(\zeta_s(0,t)-1))dt}_{\text{stress-free boundary condition}}\\ +\underbrace{\int_{0}^{T}\theta_2(t)(m(L,t)+k(\zeta_s(L,t)-1))dt.}_{\text{stress-free boundary condition}}
\end{multline}
Here $\alpha$, $\lambda(s,t)$, $\theta_1(t)$ and $\theta_2(t)$ are Lagrange multipliers.  To make this problem analytically tractable, we expand each term as a perturbation series and rewrite it using the corresponding non-zero leading order terms. 

\subsubsection{COM displacement}
From \eqref{eq:COMvel}, the COM displacement is given by 
\begin{equation*}
    \delta\zeta^{com} =\int_{0}^{T}\zeta^{com}_{t}(t)dt = \int_{0}^{T}\sum_{i=1}^{\infty}\chi^i<\beta^i_t(s,t')>_Ldt'=\frac{1}{L}\int_{0}^{T}\int_{0}^{L}\sum_{i=1}^{\infty}\chi^i\beta^i_t(s',t')dt'ds'.
\end{equation*}
We showed that the zeroth-order term and the spatial average of $\beta^1_t(s,t)$ are zero. Therefore the leading order non-zero dynamics is given by the $O(\chi^2)$ term. We use (\ref{eq:O2Relation}) to write $<\beta^2_t(s,t)>_{L}$ in terms of first-order variables as
\begin{equation}\label{eq:SecondOrderGeneralzetaCom}
    \delta\zeta^{com} = \frac{\chi^2}{L}\int_{0}^{T}\int_{0}^{L} \beta^{2}_t(s,t)ds dt+O(\chi^3)=\frac{-a\chi^2}{L}\int_{0}^{T}\int_{0}^{L}\beta^1_t(s,t)\beta^1_s(s,t)ds dt+O(\chi^3).
\end{equation}
We are thus interested in maximizing $\frac{-a}{L}\int_{0}^{T}\int_{0}^{L}\beta^1_t(s',t')\beta^1_s(s',t')ds' dt'$ and we ignore the higher order contributions.

\subsubsection{Energy constraint}
We need to constrain the energy input, which we assume is transferred to the system through the application of active stress. Therefore the energy input into the system can be written as the work done by the active stress
\begin{equation*}
    \text{Work Done by }m(s,t) = \int_{0}^{T}\int_{0}^{L}m_s(s,t)\zeta_t(s,t)dsdt= V,
\end{equation*}
where $V$ is the total energy available. We series expand $m_s(s,t)$ and $\zeta_t(s,t)$ and take the lowest order contributions, and write $V$ as perturbation series to obtain
\begin{equation*}
    \chi^2\int_{0}^{T}\int_{0}^{L}\Gamma^1_s(s,t)\beta^1_t(s,t)dsdt+O(\chi^3) = \sum_{i=0}^{i=\infty}\chi^iV_i.
\end{equation*}
We therefore constraint the equation to the lowest non-zero order given by
\begin{equation}\label{eq:PerturbEnergyConstraint}
    \int_{0}^{T}\int_{0}^{L}\Gamma^1_s(s,t)\beta^1_t(s,t)dsdt = V_2= H.
\end{equation}

\subsubsection{EOM constraint}
$\delta\zeta^{com}$ and energy constraint involve integrals in $\beta^1_t(s,t)$, $\beta^1_s(s,t)$, $\Gamma^1_s(s,t)$ which are not independent but are connected by the EOM in $O(\chi^1)$ 
\begin{equation}\label{eq:O1EOM}
    -\gamma \beta^1_{t}(s,t)+\Gamma^1_s(s,t)+k\beta^1_{ss}(s,t)=0.
\end{equation}
We encode this dependence as a constraint using a Lagrange multiplier.

\subsubsection{Stress-free boundary conditions}
Since the boundaries of the body are free, we have assumed stress-free boundary conditions (\ref{eq:SFBCi}). This constraints the boundary dynamics to 
\begin{equation*}
    \Gamma^1(0,t)+k\beta^1_s(0,t) = \Gamma^1(L,t)+k\beta^1_s(L,t) = 0.
\end{equation*}
We constraint only $O(\chi^1)$ in the stress-free boundary conditions (\ref{eq:SFBCi}) since $\delta\zeta^{com}$ and energy constraint involve integrals in $\beta^1_t(s,t)$,  $\beta^1_s(s,t)$,  $\Gamma^1_s(s,t)$ (\ref{eq:SecondOrderGeneralzetaCom}, \ref{eq:PerturbEnergyConstraint}). 

\subsubsection{Augmented functional using perturbation series quantities}
The augmented functional $S = S(\beta^1_t,\beta^1_s,\beta^1_{ss},\Gamma^1_s)$ at the first order perturbation terms is given by

\begin{multline}\label{eqn:TotalIntegralVariation}
    S = \underbrace{\frac{-a}{L}\int_{0}^{T}\int_{0}^{L}\beta^1_t(s,t)\beta^1_s(s,t)dsdt}_{\delta\zeta^{com}}+\underbrace{\alpha(\int_{0}^{T}\int_{0}^{L}\Gamma^1_s(s,t)\beta^1_t(s,t)dsdt - H)}_{\text{energy input constraint}} \\ +\underbrace{\int_{0}^{T}\int_{0}^{L}\lambda(s,t)(-\gamma \beta^1_{t}(s,t)+\Gamma^1_s(s,t)+k\beta^1_{ss}(s,t))dsdt}_{\text{EOM}}+\underbrace{\int_{0}^{T}\theta_1(t)(\Gamma^1(0,t)+k\beta^1_s(0,t))dt}_{\text{stress-free boundary condition}} \\
    +\underbrace{\int_{0}^{T}\theta_2(t)(\Gamma^1(L,t)+k\beta^1_s(L,t))dt.}_{\text{stress-free boundary condition}} 
\end{multline}

\subsection{Solving the Variational Problem}\label{sec:SolvingVariationalProblem}
\subsubsection{Extremum Equations}
The augmented integrand of (\ref{eqn:TotalIntegralVariation}) is given by
\begin{equation*}
    \tilde{L}(\beta^1_t,\beta^1_s,\beta^1_{ss},\Gamma^1_s) = \frac{-a}{L}\beta^1_t(s,t)\beta^1_s(s,t)+\alpha\Gamma^1_s(s,t)\zeta_t(s,t)+\lambda(s,t)(-\gamma \beta^1_{t}(s,t)+\Gamma^1_s(s,t)+k\beta^1_{ss}(s,t)).
\end{equation*}
To find the integral extremizers, we seek the solutions that satisfy $\delta S = 0$, which can be computed from the Euler-Lagrange equations \cite{CalcOfVariations}
\begin{equation}\label{eq:betaextr}
     -\frac{d}{ds}\frac{\partial \tilde{L}}{\partial \beta^1_s}-\frac{d}{dt}\frac{\partial \tilde{L}}{\partial \beta^1_t}+\frac{d^2}{ds^2}\frac{\partial \tilde{L}}{\partial \beta^1_{ss}}=0\rightarrow \frac{2a\beta^1_{ts}}{L}-\alpha\Gamma^1_{ts}-\gamma\lambda_t-k\lambda_{ss}=0,
\end{equation}
\begin{equation}\label{eq:GammaExtr}
     -\frac{d}{ds}\frac{\partial L}{ \partial \Gamma^1_s} =0 \rightarrow \alpha \beta^1_{ts}-\lambda_s=0.
\end{equation}
The natural boundary conditions of the variational problem accounts for the last two terms in \eqref{eqn:TotalIntegralVariation}, and are given by
\begin{equation}\label{eq:NBCs0}
    \frac{\partial \tilde{L}}{\partial \beta^1_{ss}} = -k\theta_1(t) \quad \frac{\partial \tilde{L}}{\partial \Gamma^1_{s}} = -\theta_1(t)\rightarrow \frac{\partial \tilde{L}}{\partial \Gamma^1_{s}} =\frac{1}{k}\frac{\partial \tilde{L}}{\partial \beta^1_{ss}}\Rightarrow \beta^1_t(0,t)= 0\text{ }\forall t,
\end{equation}
\begin{equation}\label{eq:NBCsL}
    \frac{\partial \tilde{L}}{\partial \beta^1_{ss}} = -k\theta_2(t) \quad  \frac{\partial \tilde{L}}{\partial \Gamma^1_{s}} = -\theta_2(t)\rightarrow \frac{\partial \tilde{L}}{\partial \Gamma^1_{s}} =\frac{1}{k}\frac{\partial \tilde{L}}{\partial \beta^1_{ss}} \Rightarrow \beta^1_t(L,t)= 0\text{ }\forall t,
\end{equation}

\begin{equation*}
    \frac{d}{ds}\frac{\partial \tilde{L}}{\partial \beta^1_{ss}} = \frac{\partial \tilde{L}}{\partial \beta^1_{s}} \Rightarrow k\lambda_s(L,t) = \frac{a}{L}\beta^1_{t}(L,t) \text{ }\forall t,
\end{equation*}

\begin{equation*}
     \frac{d}{ds}\frac{\partial \tilde{L}}{\partial \beta^1_{ss}} = \frac{\partial \tilde{L}}{\partial \beta^1_{s}} \Rightarrow k\lambda_s(L,t) = \frac{a}{L}\beta^1_{t}(L,t)\text{ }\forall t.
\end{equation*}

\subsubsection{Solving the Extremum equations}
Solving (\ref{eq:GammaExtr}) we obtain 
\begin{equation}\label{eq:beta1inlamb}
    \beta^1_t(s,t) = \frac{\lambda(s,t)+g(t)}{\alpha}.
\end{equation}
Using (\ref{eq:O1EOM}) and (\ref{eq:beta1inlamb}), we have 
\begin{equation}\label{eq:gammats}
    \alpha \Gamma^1_{ts}(s,t) = \gamma\lambda_t(s,t)+ \gamma g_t(t)-k\lambda_{ss}(s,t).
\end{equation}
Substituting (\ref{eq:gammats}) in (\ref{eq:betaextr}) we get,
\begin{equation*}
    \lambda_t(t)+\frac{g_t(t)}{2}=\frac{a}{L\gamma\alpha}\lambda_s(s,t)\rightarrow\tilde{\lambda}_t(s,t)=\frac{a}{L\gamma\alpha}\tilde{\lambda}_s(s,t);\text{ }\tilde{\lambda}(s,t) = \lambda(s,t)+\frac{g(t)}{2}.
\end{equation*}
Since $\tilde{\lambda}(s,t)$ satisfies a first-order wave equation, we can write the solution of $\lambda(s,t) = \frac{-g(t)}{2}+f(s+ct)$ where $c=\frac{a}{L\gamma\alpha}$. The shape of the travelling wave solution of $\lambda(s,t)$ is given by $f(x)$ . Substituting the functional form of $\lambda(s,t)$ in $\beta^1_t(s,t)$ we get,
\begin{equation}\label{eq:BetaInG}
    \beta^1_t(s,t) = \frac{f(s+ct)+\frac{g(t)}{2}}{\alpha}.
\end{equation}
Since $<\beta^1_t(s,t)>_L=0$, we obtain
\begin{equation}\label{eq:EqnG}
    \frac{g(t)}{2} = -< f(s+ct)>_L.
\end{equation}
Substituting (\ref{eq:EqnG}) in (\ref{eq:BetaInG}) and assuming that the crawler starts from an unstretched configuration, we obtain 
\begin{equation}\label{eq:beta1tfinalfrm}
    \beta^1_t(s,t) = \frac{f(s+ct)-<f(s+ct)>_L}{\alpha},
\end{equation}
\begin{equation}\label{eq:beta1sfinalfrm}
    \beta^1_s(s,t) = \frac{f(ct+s)-f(s)}{v\alpha}.
\end{equation}
We substitute (\ref{eq:beta1tfinalfrm}) and (\ref{eq:beta1sfinalfrm}) in (\ref{eq:O1EOM}) to solve for $\Gamma^1(s,t)$ and obtain 
\begin{equation*}
    \Gamma^1(s,t) =\frac{\gamma}{\alpha}(\int_0^sf(s'+ct)ds'-\frac{s}{L}\int_0^Lf(s'+ct)ds')-\frac{k}{c\alpha}(f(ct+s)-f(ct)-(f(s)-f(0)))+\Gamma(0,t).
\end{equation*}
We further use (\ref{eq:SFBCi}) to simplify $\Gamma^1(s,t)$ and obtain the optimal active stress distribution 
\begin{equation}\label{eq:Finalmst}
    m(s,t) = \chi(\frac{\gamma}{\alpha}(\int_0^sf(s'+ct)ds'-\frac{s}{L}\int_0^Lf(s'+ct)ds')-\frac{k}{c\alpha}(f(ct+s)-f(s)))+O(\chi^2).
\end{equation}

\subsection{Spectral Analysis of the extremizing solution}\label{sec:SpectralAnalysis}
To analyze the optimal solution obtained, we write the variables in terms of Fourier series. We can write $\beta^1_t(s,t)$ as a Fourier series in  $s\in [0,L]$
\begin{equation}\label{eqn:beta1tfourierseries}
    \beta^1_t(s,t) = a_0(t) +\sum_{n=1}^{n=\infty}a_n(t)\sin(q_ns)+b_n(t)\cos(q_ns) \text{,\  }q_n = \frac{2\pi n}{L}.
\end{equation}
 Since $<\beta_t^1(s,t)>_L = 0 $, we have $a_0(t)=0$. We evaluate both (\ref{eq:beta1tfinalfrm}) and (\ref{eqn:beta1tfourierseries}) at $ t=0 $ to find the Fourier series expansion of $f(s)$
 \begin{equation*}
     f(s)  = <f(s)>_L+\alpha\sum_{n=1}^{n=\infty}a_n(0)\sin(q_ns)+b_n(0)\cos(q_ns).
 \end{equation*}
In the following equations, we use $a_n = a_n(0)$ and $b_n = b_n(0)$. The Fourier series expansions for $\beta_t^1(s,t),\beta_s^1(s,t) $ and $\Gamma^1(s,t)$ are given by
\begin{equation}\label{eqn:beta1tfourierfinal}
    \beta_t^1(s,t) = \sum_{n=1}^{n=\infty}a_n\sin(q_ns+q_nct)+b_n\cos(q_ns+q_nct),
\end{equation}
\begin{equation}\label{eqn:beta1sfourierfinal}
    \beta_s^1(s,t) = \frac{1}{c}(\sum_{n=1}^{n=\infty} a_n(\sin(q_n(s+ct))-\sin(q_ns))+b_n(\cos(q_n(s+ct))-\cos(q_ns)),
\end{equation}
\begin{multline}\label{eqn:gamma1fourierfinal}
    \Gamma^1(s,t) = \gamma(\sum_{n=1}^{n=\infty}a_n\frac{(\cos(q_nct)-\cos(q_n(s+ct)))}{q_n}+b_n\frac{(\sin(q_n(s+ct))-\sin(q_nct))}{q_n})\\-\frac{k}{c}(\sum_{n=1}^{n=\infty}a_n(\sin(q_n(s+ct))-\sin(q_ns))+b_n(\cos(q_n(s+ct))-\cos(q_ns))).
\end{multline}
We define the initial kinetic energy $KE_{initial} = \Delta_0+\chi\Delta_1+\chi^2\Delta_2+O(\chi^3)$, which can be computed as 
\begin{equation*}
    KE_{initial} = \int_0^L\frac{\rho_0}{2}\zeta_t^2(s,0)ds = \chi^2\frac{1}{2L}\int_0^L (\beta^1_t(s,0))^2ds+O(\chi^3)\approx \chi^2\frac{1}{4}\sum_{n=1}^{n=\infty}(a_n^2+b_n^2)+O(\chi^3).
\end{equation*}
We consider the lowest non-zero order for our analysis, therefore we get $\Delta_0=0$, $\Delta_1=0$ and $\Delta_2 = \frac{1}{4}\sum_{n=1}^{n=\infty}(a_n^2+b_n^2)$. We can now use this representation to calculate $\delta \zeta^{com}(t)$ (\ref{eq:SecondOrderGeneralzetaCom}) and the total energy input (\ref{eq:PerturbEnergyConstraint}) into the system as follows 

\begin{equation}\label{eqn:deltaxfourier}
    \delta \zeta^{com}(t) = \frac{-a}{L}\int_{0}^{t}\int_{0}^{L}\beta^1_t(s,t)\beta^1_s(s,t)ds dt = -2L\gamma\alpha t\Delta_2+\frac{L^3\gamma^2\alpha^2}{4\pi a}\sum_{n=1}^{n=\infty}\sin(\frac{2n\pi ta}{L^2\gamma\alpha})(\frac{a_n^2+b_n^2}{n}),
\end{equation}
\begin{equation}\label{eqn:energyconstraintfourier}
    H = \int_{0}^{T}\int_{0}^{L}\Gamma^1_s(s,t)\beta^1_t(s,t)ds dt =2L\gamma T\Delta_2+\frac{2L^3\gamma^2k\alpha^2}{a^2}(\Delta_2-\sum_{n=1}^{n=\infty}\cos(\frac{2n\pi Ta}{L^2\gamma\alpha})\frac{(a_n^2+b_n^2)}{4}).
\end{equation}

\subsection{Comparing perturbation series solutions with complete nonlinear equation}

\begin{figure}[h]
\centering
\includegraphics[width=0.7\textwidth]{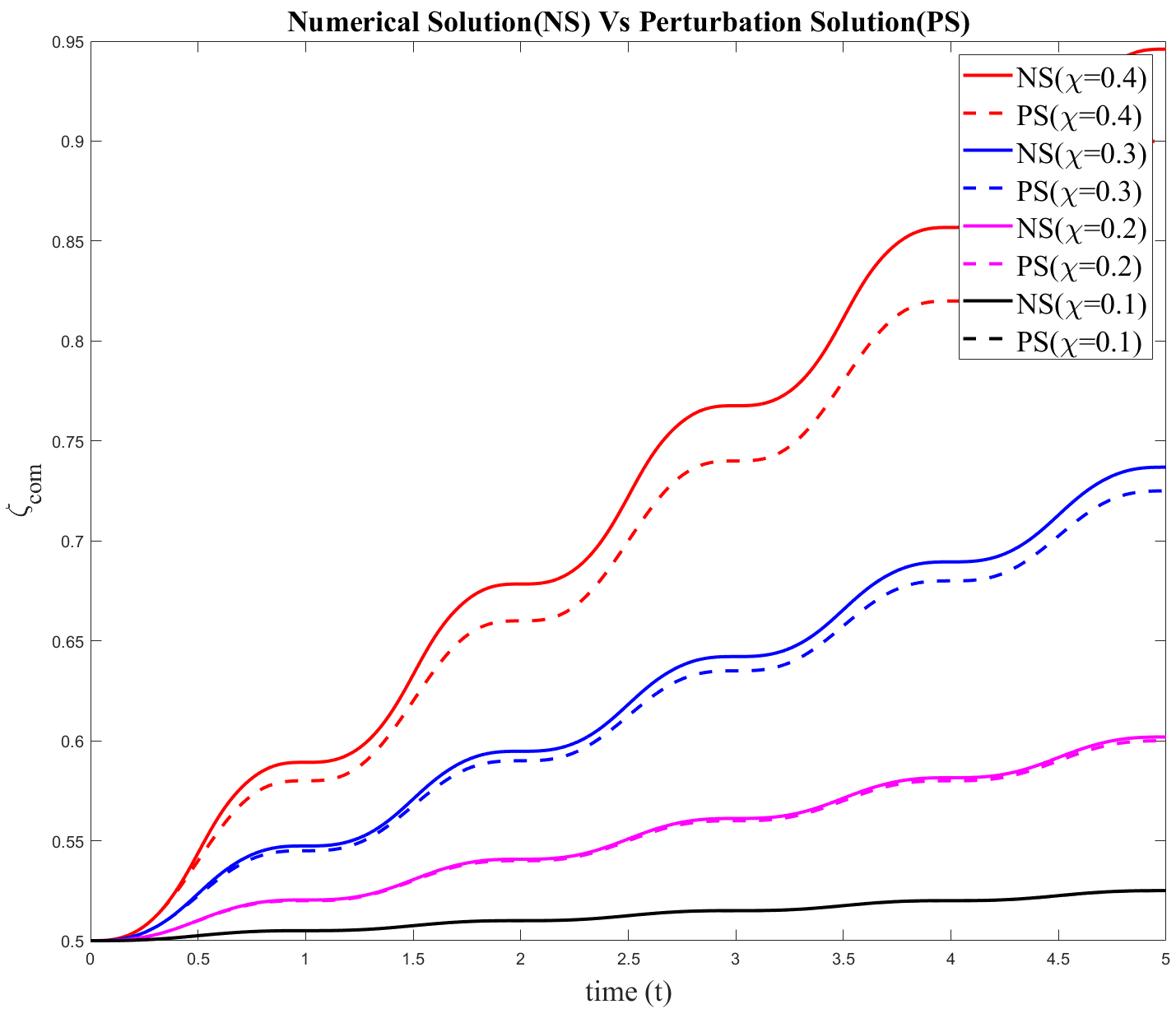}
\caption{Comparison of numerical solution of COM displacement using the complete nonlinear \textcolor{black}{EOM} (\ref{eq:FinalLagEOM}) and the perturbation solution (PS) (\ref{eqn:deltaxfourier}) for different values of $\chi$. The numerical solution and the perturbation solution converges as $\chi$ becomes smaller.}
\label{fig:zetaCOM}
\end{figure}

We check the validity of our perturbation series solution by comparing the motion of the center of mass with the one of the full nonlinear system for various values of perturbation parameter $\chi$. Specifically, we compare the numerical result of the complete nonlinear EOM (\ref{eq:FinalLagEOM}) with the analytical result derived from perturbation theory (\ref{eqn:deltaxfourier}). Fig. \ref{fig:zetaCOM} shows the validity of our analytical prediction and its increasing accuracy for decreasing $\chi$. In our analysis, we used the following set of parameters, $a=-1,\text{ }\gamma=1,\text{ }k=1,\text{ }H= 0.5,\text{ }L=1,\text{ }T=1,\text{ }a_1 = 1,\text{ }a_{n\neq1}=0,\text{ }b_n = 0 \text{ }\forall \text{ }n\in \mathbb{N}^+, n>0.$

\subsection{Scaling relations}\label{sec:ScalingRelations}

We use \eqref{eq:xComTheorem} and \eqref{eq:EnergyAlpha} to study the scaling of the energy-efficient crawling gait given by Theorem \ref{theorem} with the body mass $M$. Using \eqref{eq:xComTheorem}, and substituting $\alpha = \frac{a}{L\gamma c}$, the average crawling speed $U$ normalised to the body length $L$ across a time period $T$ of the gait is given by
\begin{equation}\label{eq:ScalingCrawlingSpeed}
    U =\frac{\delta\zeta^{com}}{LT}= -\frac{2acA^2}{L},\quad A^2 = \frac{\chi^2\Delta_2}{c^2}.
\end{equation}
Using \eqref{eq:StrainCrawler}, we see that $A$ scales with the strain amplitudes 
$\frac{\chi a_n}{c}$ and $\frac{\chi b_n}{c}$. Therefore, the average crawling speed scaling arises from the scaling of the strain amplitudes in the crawler and the scaling in the wave speed $c$. \textcolor{black}{ Here we assume that the modulation of friction by strain \eqref{eq:friction mod} is independent of the body size, therefore $a$ does not scale with the body size.}
\\
\\
A relevant quantity used to compare the energy cost of a gait across a wide range of body sizes is the cost of transport (COT) \cite{alexander1992exploring}. COT is the energy required to move a unit mass per unit distance. For the crawler gait described by Theorem \ref{theorem}, COT is given by 
\begin{equation}
    \text{COT} = \frac{V}{M\delta\zeta^{com}}.
\end{equation}
The total energetic cost during a time period $T$ \eqref{eq:EnergyAlpha} is given by 
\begin{equation}\label{eq:EnergyCostTotalFric}
    V = 2L\gamma c^2T(\frac{\chi^2\Delta}{c^2})=2L\gamma c^2\textcolor{black}{T}A^2.
\end{equation}
From \eqref{eq:EnergyCostTotalFric} and \eqref{eq:xComTheorem}, the COT is given by
\begin{equation}\label{eq:COT}
    \text{COT}= \frac{-L\gamma c}{Ma}.
\end{equation}
Since we assume an elastic response from the crawler body, the strain amplitude is constrained by the maximum tensile stress that needs to be lower than the failure stress, independent of the body size and dependent on the body material. Therefore we have $A \propto M^{0}$. We assume that the crawler grows isometrically. Therefore all length dimensions scale as $M^{\frac{1}{3}}$ \cite{alexander1992exploring}. The frictional coefficient per unit length  scales with the width of the surface of the crawler in touch with the substrate, i.e. $\gamma\propto M^{\frac{1}{3}}$. Finally, we write the wave speed $c = L\nu$, where $\nu$ is the stride frequency. From the above arguments, $U$ scales as 
\begin{equation}
    U \propto \nu,
\end{equation}
and the COT as
\begin{equation}
    \text{COT} \propto \nu.
\end{equation}
\newpage
\normalem
\bibliographystyle{ieeetr}
\bibliography{bibNew}

\end{document}